# Superconductivity and phase diagram in Sr-doped $La_3Ni_2O_7$ thin films

*Bo Hao[1, 2, ‡], Maosen Wang[1, 2, ‡], Wenjie Sun[1, 2, ‡], Yang Yang[3, ‡], Zhangwen Mao[1, 2], Shengjun Yan[1, 2], Haoying Sun[1, 2], Hongyi Zhang[1, 2], Lu Han[1, 2], Zhengbin Gu[1, 2], Jian Zhou[1, 2], Dianxiang Ji[3, *], Yuefeng Nie[1, 2, 4, *]*

[1]National Laboratory of Solid State Microstructures, Jiangsu Key Laboratory of Artificial Functional Materials, College of Engineering and Applied Sciences, Nanjing University, Nanjing 210093, P. R. China.

[2]Collaborative Innovation Center of Advanced Microstructures, Nanjing University, Nanjing 210093, P. R. China.

[3]Department of Applied Physics, The Hong Kong Polytechnic University, Hung Hom, Hong Kong, China

[4]Jiangsu Physical Science Research Center, Nanjing 210093, China.

‡ These authors contributed equally to this work.
* Corresponding author: Dianxiang Ji, Yuefeng Nie

## Abstract

Recent studies have demonstrated ambient pressure superconductivity in compressively strained $La_3Ni_2O_7$ thin films, yet the phase diagram of heterovalent doping—critical for advancing the field—remains unexplored. Here, we report superconductivity in $Sr^{2+}$-doped $La_{3-x}Sr_xNi_2O_7$ films synthesized via molecular beam epitaxy with ozone-assisted post-annealing. The superconducting transition temperature ($T_c$) follows an asymmetric dome-like profile, persisting across a wide doping range ($0 \leqslant x \leqslant 0.21$) before diminishing at $x \approx 0.38$. Optimally doped films ($x = 0.09$) achieve $T_c$ of ~ 42 K, with high critical current ($J_c > 1.4$ kA/cm² at 2 K) and upper critical fields ($\mu_0 H_{c,\parallel}(0) = 83.7$ T, $\mu_0 H_{c,\perp}(0) = 110.3$ T), comparable to reported $La_{3-x}Pr_xNi_2O_7$ films. Scanning transmission electron microscopy reveals oxygen vacancies predominantly occupy at planar $NiO_2$ sites—unlike apical-site vacancies in bulk samples—due to Coulomb repulsion destabilizing planar oxygen under compressive strain. Additionally, the elongated out-of-plane Ni–O bonds, exceeding those in pressurized bulk samples by 4%, likely weaken the interlayer $d_{z^2}$ coupling, thus contributing to the reduced $T_c$ in strained films. This work establishes heterovalent $Sr^{2+}$ doping as a robust tuning parameter for nickelate superconductivity, unveiling a unique phase diagram topology.

## Main

The discovery of high-temperature superconductivity in Ruddlesden-Popper (RP) nickelates ($A_{n+1}Ni_nO_{3n+1}$), with a superconducting transition temperature near 80 K under high pressure, has arisen significant attention.[1-13] However, achieving high-$T_c$ superconductivity in bulk nickelates requires high-pressure conditions, hindering the characterizations critical for understanding the electronic structure and pairing symmetry of nickelate superconductors. To address this limitation, recent studies have demonstrated that the compressive epitaxial strain can stabilize superconductivity in $La_3Ni_2O_7$ films epitaxially grown on $SrLaAlO_4$ substrates[14]. Meanwhile, partial substitution of La with Pr in $La_{3-x}Pr_xNi_2O_7$ films has been shown to enhance crystallinity and superconductivity, with the onset transition temperature ($T_{c,onset}$) reaching 48 K[15,16]. Similar enhancement has been previously observed in bulk samples through substitution with Pr and Sm[5,17-20], where the improvement is primarily attributed to the suppression of intergrown RP phases.

In comparison to the abovementioned isovalent substitution, which preserves the electronic configuration of Ni, heterovalent doping offers a powerful knob for tuning the electronic states of quantum materials. In high-$T_c$ cuprates, superconductivity emerges through carrier doping, which suppresses the antiferromagnetic parent state, and gives rise to a complex phase diagram encompassing pseudogap, strange metal, and Fermi-liquid regimes[21-25]. This paradigm highlights the critical role of carrier concentration in optimizing superconductivity, yet its applicability to the emerging nickelate superconductors remains an open frontier. Bilayer nickelates exhibit a distinct multiband electronic structure consisting of La 5$d$ orbitals and Ni 3$d$ orbitals, with the metallization of $3d_{z^2}$ orbital under high pressure proposed to play an important role in superconductivity[1,26-32]. In thin films, theoretical studies indicate that compressive strain shifts the $d_{z^2}$-band-dominated $\gamma$ band down away from the Fermi level ($E_F$)[33,34]. Hole doping, however, elevates this $\gamma$ band to cross the $E_F$[33-36], akin to that in the high-pressure bulk samples[1]. Nonetheless, experimental efforts to dope $La_3Ni_2O_7$ with alkaline-earth metals (e.g., Sr or Ca) in both bulk and thin-film forms have proven challenging, as superconductivity has not been observed to date, even under high pressure[37-39].

Here, we report the superconductivity and its dependence on Sr doping in compressively strained $La_{3-x}Sr_xNi_2O_7$ thin films, synthesized through reactive molecular-beam epitaxy (MBE) followed by ozone-assisted post-annealing. The optimal superconducting transition temperature ($T_c$) is comparable to that observed in undoped counterparts[14-16]. Through Sr doping, superconductivity with a similar $T_c$ persists up to a doping level of $x$ = 0.21, above which it gradually diminishes and vanishes by $x$ = 0.38. This results in an asymmetric phase diagram, distinct from the more symmetric superconducting dome typically observed in cuprates.

## Epitaxial film growth of Sr-doped $La_3Ni_2O_7$ films

High-quality 3-unit-cell (3 u.c.) $La_{3-x}Sr_xNi_2O_7$ thin films with Sr doping levels ($x$) ranging from 0 to 0.45 were grown on (001)-oriented $SrLaAlO_4$ (SLAO) substrates in an oxidant (distilled ozone) atmosphere using MBE. Each film was coated with 1 u.c. $SrTiO_3$ (STO) as the capping layer. The stacking sequence and thickness of the films were precisely controlled using a shuttered method. (Details of growth are discussed in Methods and **Extended Data Fig. 1**). The crystalline quality of these films was confirmed by x-ray diffraction (XRD) $2\theta$-$\omega$ scans along the (00$l$) direction (**Fig. 1a**), where all diffraction peaks align well and there is no signature of impurity phases. The Kiessig fringes, along with atomic force microscopy (AFM) results, indicate atomically smooth surface and interface. Reciprocal space mapping (RSM) around the $SrLaAlO_4$ ($10\underline{11}$) and $La_{2.79}Sr_{0.21}Ni_2O_7$ ($10\underline{17}$) reflections (**Fig. 1b**) confirms that the film is coherently strained to the substrate, with an in-plane lattice constant of 3.75 Å, identical across all samples with varying Sr doping levels. The $c$-axis lattice constant of the undoped $La_3Ni_2O_7$ film on the SLAO substrate is measured as 20.85 Å, representing an elongation of approximately 1.6% compared to the bulk value. Upon Sr doping, the $c$-axis lattice constant decreases, reaching 20.73 Å at $x = 0.45$ (**Extended Data Fig. 2**), a trend similar to that observed in bulk polycrystalline $La_3Ni_2O_7$[37]. Furthermore, the full width at half maximum (FWHM) of rocking curve (**Fig. 1c**) for $La_{2.79}Sr_{0.21}Ni_2O_7$ film is comparable to that of $SrLaAlO_4$ substrate (about 0.024° for both), indicating the high crystalline quality of the films.

## Compressive strain effects on oxygen distribution and bond elongation

The crystallinity and interfacial integrity were confirmed by scanning transmission electron microscopy (STEM) characterizations. A large field-of-view high-angle annular dark-field (HAADF) image reveals a pure $n = 2$ RP structure, without evidence of intergrown RP phases (**Fig. 2a**). Atomic-resolution energy-dispersive X-ray spectroscopy (EDS) elemental maps demonstrate a sharp interface between the $La_{3-x}Sr_xNi_2O_7$ film and the SLAO substrate, without detectable ionic interdiffusion (**Fig. 2d**). Furthermore, Sr is uniformly distributed within the film with negligible Sr diffusion from the substrate, different from previous report of Sr segregation near the interface from the substrate[15]. As noted in the previous report[15], a surface reconstruction (the area between the dashed lines) is also observed on the SLAO substrate in our work, likely formed during pre-growth annealing[40].

Integrated differential phase contrast (iDPC) imaging was employed to reveal the detailed information about the oxygen location and concentration. As shown in **Fig. 2b**, the iDPC images exhibit two notable observations:

1) Site-dependent oxygen intensity variations and oxygen vacancies: The oxygen intensity correlates strongly with positions. Apical oxygen atoms (inner and outer sites) exhibit enhanced intensity compared to the planar oxygen atoms within the $NiO_2$ layers (**Fig. 2c**). This variation in intensity suggests a distribution of oxygen vacancies, likely introduced during the focused ion beam (FIB) process used for STEM sample preparation in a vacuum environment. Such observations imply that oxygen vacancies preferentially occupy the $NiO_2$ basal planes, strikingly contrasting with prior studies that reported vacancies at the apical sites in pressurized bulk counterpart[41]. In fact, this can be understood using a simplified picture. Under compressive epitaxial strain, reduced in-plane lattice constants together with elongated out-of-plane lattice constants enhances in-plane electron wavefunction overlap, increasing Coulomb repulsion at the planar sites. This elevates the energy penalty for oxygen occupancy, driving vacancy aggregation at the planar $NiO_2$ layers. DFT calculations corroborate this trend: Without external strain, the inner apical oxygen vacancy exhibits the lowest formation energy; while planar vacancy reveals its thermodynamic preference under compressive in-plane biaxial strain (**Extended Data Table 1**).

2) Ni-O bond length and Ni-O-Ni bond angle under compressive strain: Consistent with

expectations, compressive strain reduces the in-plane Ni-O bond length and elongates the out-of-plane Ni-O bond length. The out-of-plane Ni-O bond length is 1.99 Å, which is about 4% longer than that in pressurized bulk samples with $T_c$ of ~80 K[1,42]. The significant elongation of out-of-plane Ni-O bond length implies a weakened interlayer $d_{z^2}$ orbital coupling that may contribute to the suppressed $T_c$ in thin films. This may explain why thin film samples have a lower $T_c$ compared to pressurized bulk samples. Detailed lattice constants derived from STEM measurements are provided in **Extended Data Fig. 3**.

Similar results are also observed in undoped $La_3Ni_2O_7$ films (**Extended Data Fig. 4**), indicating that these structural features are predominantly influenced by compressive strain rather than Sr doping.

## Transport properties of $La_{3-x}Sr_xNi_2O_7$ films

Next, we explore the transport properties of these $La_{3-x}Sr_xNi_2O_7$ thin films. All the as-grown films were ex-situ annealed using a home-made ozone annealing system (Methods), following the strategy reported in a recent study[16]. After optimizing the annealing conditions (**Extended Data Fig. 5**), superconductivity with zero resistance is achieved in our films.

Before addressing the Sr doping dependence of $T_c$ and the phase diagram, we first examine the general transport properties of a representative sample, $La_{2.91}Sr_{0.09}Ni_2O_7$, which exhibits a relatively sharp superconducting transition (**Fig. 1d**). The onset temperature is 42 K, defined as the temperature at which the resistance drops to 98% of the extrapolated value from a linear fit to the normal state *R*-*T* data between 50 K and 60 K. This criterion is validated by comparing $\rho(T)$ curves under 0 T and 9 T magnetic fields, which showed a characteristic suppression of superconductivity under applied out-of-plane magnetic field. The zero resistance temperature ($T_{c,zero}$) reached around 22 K. Notably, the absence of a two-step resistive drop, which was observed in previous studies[14,15], indicates an improved crystalline quality and reduced disorder in our samples.

In the normal state, the temperature-dependent resistivity of $La_{2.91}Sr_{0.09}Ni_2O_7$ film deviates from the linear relation observed in bulk high-pressure samples[3]. Analysis of the resistivity derivative ($d\rho/dT$) reveals three distinct regimes: (i) a temperature-independent $d\rho/dT$ between 160 K and 200 K, (ii) a gradual decrease below 160 K resembling trends in slightly overdoped $Nd_{1-x}Sr_xNiO_2$ thin films[43,44], and (iii) stabilization between 54 K and 64 K (**Extended Data Fig. 6**). For the $\rho \sim T^n$ relation, the exponent $n$ decreases monotonically from $n \approx 2$ (65–70 K) to $n \approx 1.02$ (160–200 K). This evolution of $n$ suggests deviations from optimal doping[25,45], as inferred from comparisons with cuprate superconductors, but also could be related to residual disorder.

The parallel-resistor formula[45], previously used to fit the normal state of $La_2PrNi_2O_7$ thin films[16], successfully describes $\rho(T)$ for normal state in our films:

$$\frac{1}{\rho(T)} = \frac{1}{\alpha_0 + \alpha_1 T + \alpha_2 T^2} + \frac{1}{\rho_{sat}}$$

with fitted parameters $\alpha_0 = 0.363$ mΩ·cm, $\alpha_1 = 2.09\times10^{-4}$ mΩ·cm/K, $\alpha_2 = 1.52\times10^{-5}$ mΩ·cm/K$^2$, and $\rho_{sat} = 1.46$ mΩ·cm. Notably, the relatively small $\alpha_1$ value aligns with overdoped cuprate[45].

The Hall coefficient ($R_H$) of this sample is negative and nearly temperature-independent between 50 K and 200 K (inset in Fig. 1d), resembling the behavior observed in $La_{2.85}Pr_{0.15}Ni_2O_7$ films[15], but differing markedly from that of $La_3Ni_2O_7$[14] and $La_2PrNi_2O_7$ films[16]. These observations suggest a multiband electronic structure at the Fermi level, which aligns, in certain respects, with ARPES results on bulk crystals at ambient pressure[26]. Intriguingly, in optimally doped $La_{0.8}Sr_{0.2}NiO_2$ samples ($T_c \sim 17$ K), the Hall coefficients remain negative and attain a magnitude comparable to that of $La_{2.91}Sr_{0.09}Ni_2O_7$ just above $T_c$. (Inset in **Fig. 1d**)

Systematic magneto-transport measurements were performed on the $La_{2.91}Sr_{0.09}Ni_2O_7$ thin film under external magnetic fields up to 9 T. When the magnetic fields are applied perpendicular ($H \parallel c$) and parallel ($H \perp c$) to the *a*-*b* plane, superconductivity is gradually suppressed, and exhibits a pronounced anisotropic response (**Fig. 3a and 3b**). The upper critical fields $H_c(T)$ for both orientations are extracted using $T_{c,90\%}$ values. As illustrated in **Fig. 3c**, the out-of-plane $H_{c,\parallel}(T)$ exhibits a linear-in-temperature dependence, whereas the in-plane $H_{c,\perp}(T)$ follows a square-root dependence, and both fit the Ginzburg-Landau (GL) model well[46]. This yields zero-temperature critical fields of $\mu_0 H_{c,\parallel}(0)$ = 83.7 T and $\mu_0 H_{c,\perp}(0)$ = 110.3 T. The coherence length $\xi_{ab}(0)$ is calculated as 1.98 nm, consistent with previous reports[3,16]. The superconducting layer thickness is estimated as 5.21 nm, closely matching the 6 nm film thickness obtained from XRD fitting, suggesting that the entire film is superconducting.

To investigate the superconducting dimensionality, the angular dependence of $T_{c,50\%}$ was measured under a fixed magnetic field of 9 T (**Fig. 3d**). The $T_c(\theta)$ profile exhibits a sharp cusp-like feature near $\theta = 0°$ (H $\perp$ *c*) and is well described by the Tinkham model (solid line in **Fig. 3d**) for two-dimensional (2D) superconductors[47]. In contrast, the anisotropic Ginzburg-Landau model for 3D superconductors[48] fails to reproduce the cusp-like angular dependence, indicating the possible quasi-2D nature of the superconductivity in our films. Finally, the critical current density $J_c$ was measured as a function of temperature (**Fig. 3e**), reaching $J_c > 1.4$ kA/cm$^2$ at 2 K, which is an order of magnitude higher than those reported for $La_3Ni_2O_7$ bulk[3] and thin films[14]. A further enhancement of $J_c$ can be anticipated through partial substitution La with Pr, as suggested by isovalent doping experiments[15,16].

## Phase diagram of $La_{3-x}Sr_xNi_2O_7$

We now turn to the doping-dependent phase diagram of $La_{3-x}Sr_xNi_2O_7$ films. For low Sr doping levels ($x < 0.1$), as-grown films exhibit insulating behavior. In contrast, at intermediate doping levels ($0.1 \le x \le 0.3$), as-grown films undergo an insulator-to-superconductor transition, albeit without achieving zero resistance. After ozone annealing, normalized *R*-*T* curves reveal superconductivity with zero resistance at low doping levels and a metal-insulator transition in the high-doping regime. The ozone annealing conditions were optimized for samples across various Sr doping levels, and the highest $T_c$ values for each doping level are summarized in **Fig. 4**. The relationship between Sr doping ($x$) and superconducting transition temperatures ($T_{c,98\%}$, $T_{c,50\%}$), as depicted in the tentative phase diagram, reveals that the maximum $T_{c,onset}$ and midpoint $T_{c,50\%}$ occur at $x$ = 0.09. Starting from the superconducting undoped film, superconductivity persists across a broad doping range ($0 \leq x \leq 0.30$): $T_c$ changes slowly for $0 \leq x \leq 0.21$, then decreases monotonically and vanishes for $x > 0.30$. Samples with $x \geqslant 0.37$ exhibit metallic behavior at high temperature and weakly insulating behavior at low temperature. The suppression of superconductivity at high doping levels is unlikely due to crystalline defects, as XRD confirms comparable crystallinity across all samples (Fig. 1a). Instead, it is most likely a result of hole overdoping, similar to trends observed in cuprates[24] and infinite-layer nickelates[49,50].

## Discussion

Our study establishes heterovalent $Sr^{2+}$ doping as a robust route to engineering superconductivity in compressively strained $La_{3-x}Sr_xNi_2O_7$ thin films, achieving $T_{c,onset}$ up to 42 K under ambient pressure. The phase diagram exhibits an asymmetric profile, with superconductivity persisting across a broad doping range ($0 \leqslant x \leqslant 0.21$) before being suppressed at $x \approx 0.38$. These results contrast with the more symmetric superconducting domes observed in cuprates, likely due to the multiband electronic structure of bilayer nickelates, where hybridization between La 5*d* and Ni 3*d* orbitals mitigates carrier overdoping effects. To fully map the superconducting dome, future studies could explore $Ce^{4+}$ substitution for $La^{3+}$, systematically investigating the electron-doped regime for potential electron-hole asymmetry—a hallmark of unconventional superconductivity.

Furthermore, the dual effects of compressive strain—enhanced in-plane hopping via shortened Ni–O bonds versus weakened interlayer $d_{z^2}$ coupling due to elongated out-of-plane bonds—highlight a delicate competition governing $T_c$. The observed similar in-plane Ni-O bonds and 4% elongation of out-of-plane Ni–O bonds compared to pressurized bulk samples provides a structural rationale for the lower $T_c$ in bilayer nickelate films, suggesting that interlayer coupling plays a critical role in pairing strength. This strain-mediated competition implies the possible existence of an optimal strain window, beyond which further compression may degrade superconductivity.

On the other hand, oxygen stoichiometry is crucial for achieving superconductivity in both bulk samples and thin films, albeit through different mechanisms. In bulk samples, preferential inner-apical oxygen vacancies are believed to disrupt interlayer $d_{z^2}$ coupling and suppress superconductivity[28]. In contrast, compressive strain in thin films favors planar oxygen vacancies, but the resulting disorder scatters carriers within the superconducting plane, suppressing superconductivity similarly to overdoped cuprates. Thus, oxygen stoichiometry plays a universal role in bilayer nickelate superconductivity, requiring high-pressure oxygen annealing for bulk samples and ozone-assisted annealing for thin films to optimize superconducting properties.

Last but not least, systematic analysis of carrier concentration via Hall effect measurements remains limited by multiband complexities and sample heterogeneity across a wide range of Sr doping levels. Future efforts should prioritize improving sample homogeneity for reliable Hall characterization. Additionally, orbital-specific probes such as resonant inelastic X-ray scattering (RIXS) and angle-resolved photoemission spectroscopy (ARPES) are highly desirable to elucidate how $d_{z^2}$ and $d_{x^2-y^2}$ states evolve under strain and doping, to further clarify their roles in the pairing mechanism.

## Figures and captions

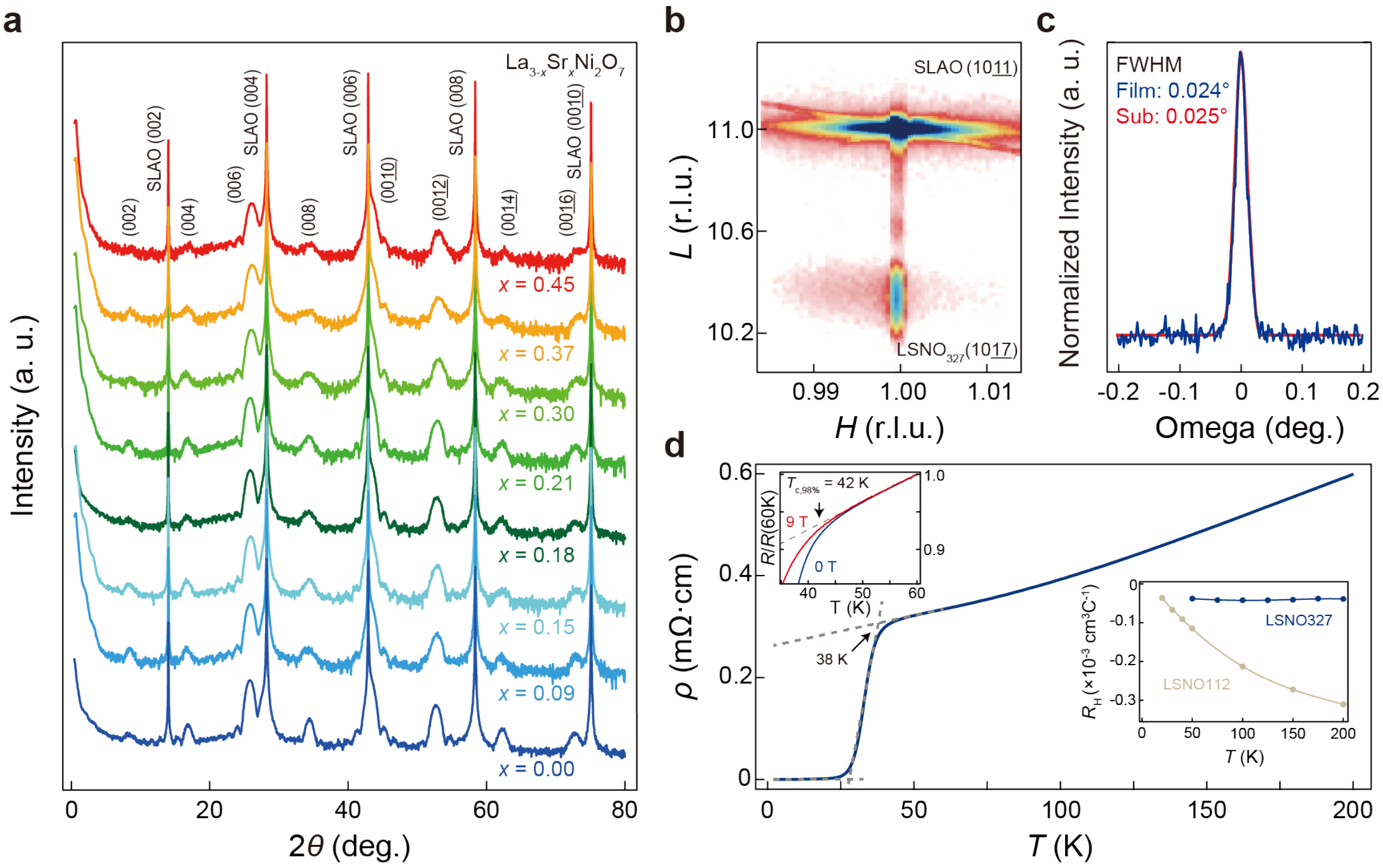

**Figure 1. Structural properties and superconductivity of $La_{3-x}Sr_xNi_2O_7$ thin films.** **a,** X-ray diffraction 2$\theta$-$\omega$ scans of $La_{3-x}Sr_xNi_2O_7$ thin films grown on $SrLaAlO_4$ (001) substrates. **b,** Reciprocal space mapping around the $SrLaAlO_4$ (10<u>11</u>) and $La_{2.79}Sr_{0.21}Ni_2O_7$ (10<u>17</u>) reflections, confirming coherent epitaxial strain between the film and substrate. **c,** Rocking curve of the $SrLaAlO_4$ (004) and $La_{2.79}Sr_{0.21}Ni_2O_7$ (006) peaks, exhibiting comparable full-width-at-half-maxima (FWHM). **d,** Temperature-dependent resistivity (200-2 K) of a $La_{2.91}Sr_{0.09}Ni_2O_7$ film. The onset $T_c$ is defined as the temperature at which the resistance equals 98% of the extrapolated value from a linear fit between 50 K and 60 K (gray dashed line), and can be confirmed by comparing with $R$-$T$ curve under 9 T magnetic field (up-left inset). Bottom-right inset shows the temperature-dependent Hall coefficients for the $La_{2.91}Sr_{0.09}Ni_2O_7$ film (blue circles), and Hall coefficients for a $La_{0.8}Sr_{0.2}NiO_2$ sample ($T_c$ = 17 K, grey circles) are also shown for comparison.

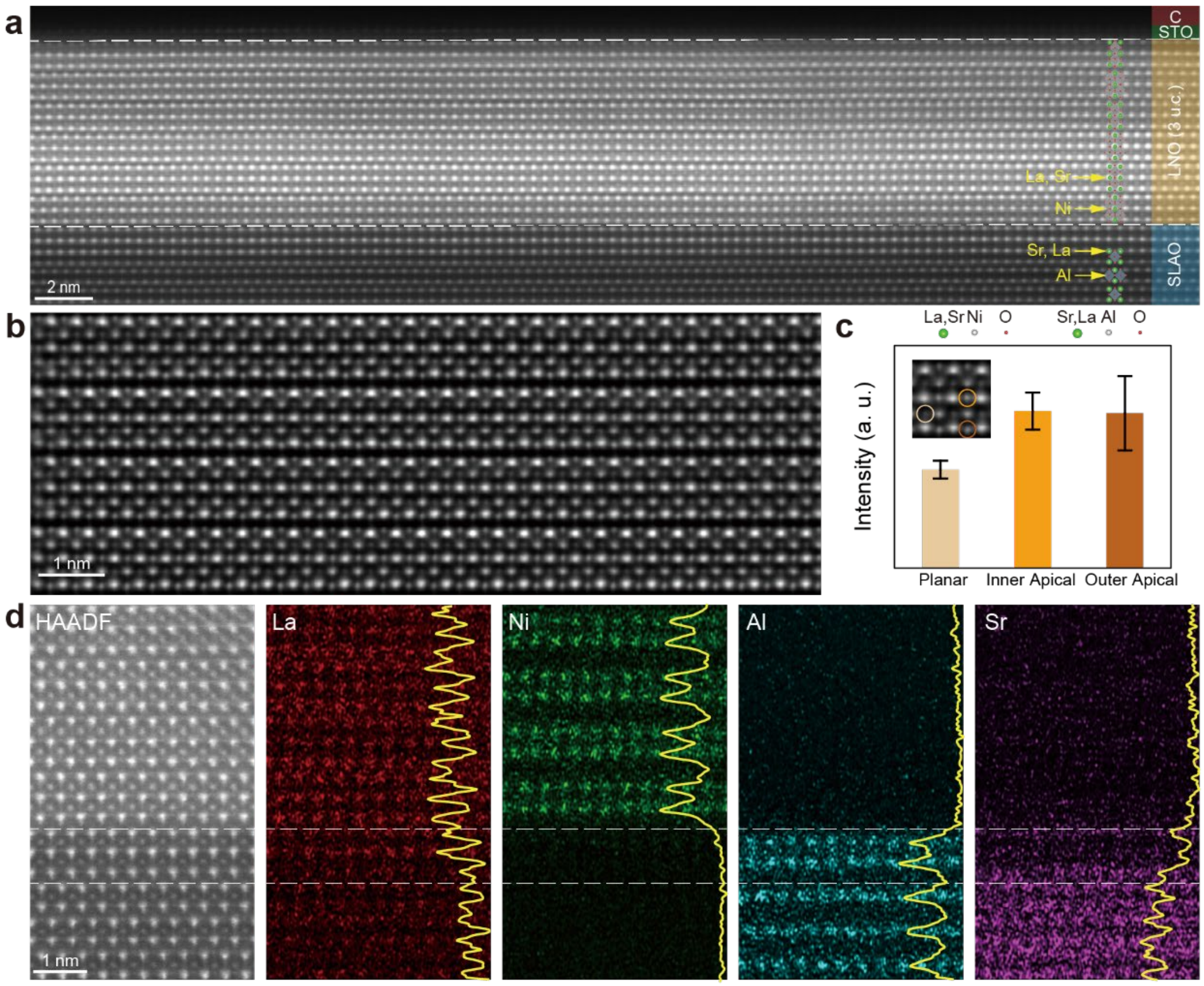

**Figure 2. Scanning transmission electron microscopy (STEM) characterizations of a superconducting $La_{2.79}Sr_{0.21}Ni_2O_7$ film. a**, High-angle annular dark-field (HAADF) image with a large field-of-view of a 3-u.c.-thick $La_{2.79}Sr_{0.21}Ni_2O_7$ film grown on a $SrLaAlO_4$ substrate. Dash lines represent the interface between the $SrTiO_3$ capping layer and the film and the interface between the film and the $SrLaAlO_4$ substrate, respectively. **b**, Integrated differential phase contrast (iDPC) image of the $La_{2.79}Sr_{0.21}Ni_2O_7$ film. **c**, The intensity histogram of oxygen column of three distinctive sites as indicated. **d**, HAADF image and atomic-resolution energy-dispersive X-ray spectroscopy (EDS) elemental maps (La, Ni, Al, Sr) of the same region. The yellow curves represent the profiles of atomic row-integrated elements intensity. The area between the dashed lines indicates the surface reconstruction region of the $SrLaAlO_4$ substrate.

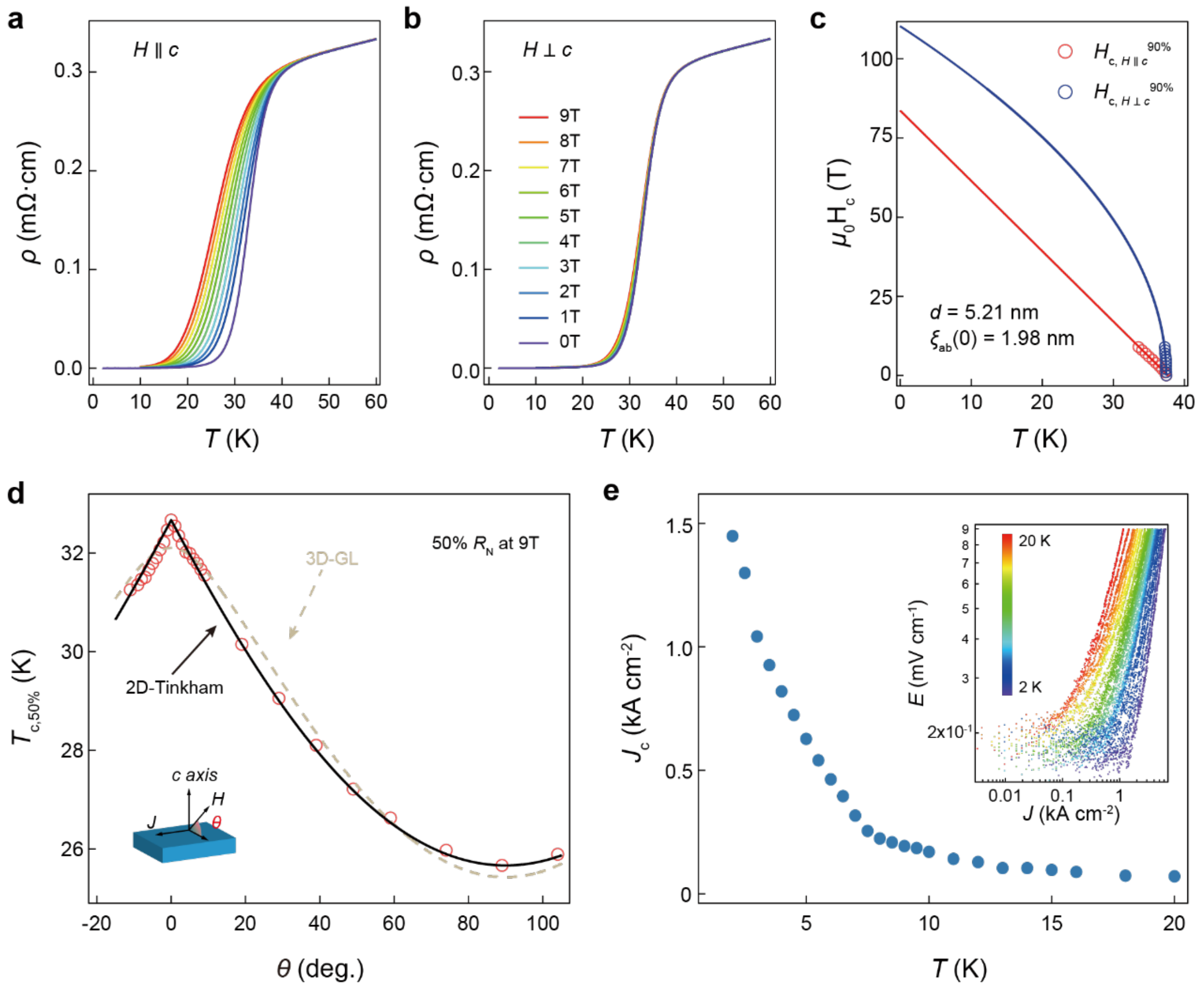

**Figure 3. Transport properties of $La_{2.91}Sr_{0.09}Ni_2O_7$ thin films. a-b,** $\rho(T)$ curves of $La_{2.91}Sr_{0.09}Ni_2O_7$ thin film under various magnetic fields applied perpendicular (**a**) and parallel (**b**) to the *a*-*b* plane of the film. **c,** Perpendicular and parallel upper critical fields extracted by the $T_{c,90\%}$ values (open circles) respectively. Solid lines represent Ginzburg-Landau fitting results. **d,** Angular dependence of $T_{c,50\%}$ at 9 T fitted with the 2D-Tinkham model (solid line) and 3D-Ginzburg-Landau model (dashed line). Open circles are the measured $T_{c,50\%}$ values. Inset shows the measurement configuration, where $\theta$ is the angle between magnetic field and the film *a*-*b* plane. **e,** Temperature dependent critical current density $J_c$. Inset shows the corresponding electric field ($E$) versus current density ($J$) curves measured between 2 to 20 K.

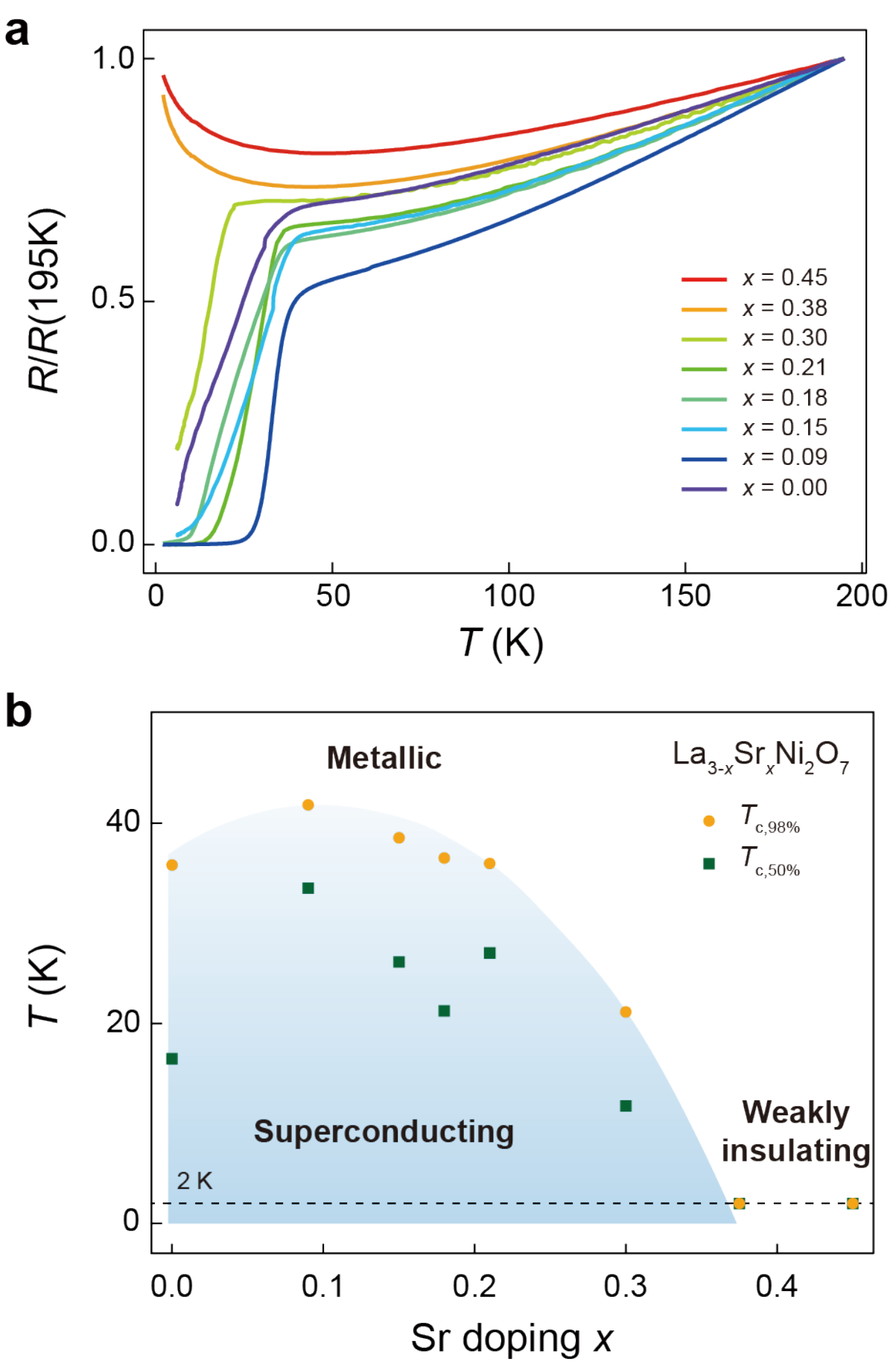

**Figure 4. Proposed phase diagram of $La_{3-x}Sr_xNi_2O_7$ thin films. a,** Temperature-dependence of resistivity (normalized at 195 K) for the representative samples with Sr doping levels $x$ = 0, 0.09, 0.15, 0.18, 0.21, 0.30, 0.38, and 0.45. **b,** Doping-dependent superconducting phase diagram of $La_{3-x}Sr_xNi_2O_7$ thin film. Circles (squares) represent $T_{c,98\%}$ ($T_{c,50\%}$), defined as the temperature at which the resistivity reaches 98% (50%) of the extrapolated normal-state resistivity linearly fitted above the superconducting transition.

## Methods

### Thin film growth

The $La_{3-x}Sr_xNi_2O_7$ thin films with STO capping layers were synthesized via reactive molecular-beam epitaxy (MBE) in a DCA R450 system on (001)-oriented $SrLaAlO_4$ substrates (supplied by HF-Kejing). During growth, the substrate temperature was maintained at 720°C under a distilled ozone background pressure of $1 \times 10^{-5}$ Torr. Flux calibration was performed as follows: First, the evaporation rates of La, Ni, Sr, and Ti sources were preliminarily calibrated using a quartz crystal microbalance (QCM) under high vacuum. Subsequently, precise flux optimization was achieved by monitoring reflection high-energy electron diffraction (RHEED) oscillations during the co-deposition of $LaNiO_3$ on SLAO and $SrTiO_3$ on $SrTiO_3$ substrates, respectively. Prior to transfer into the MBE chamber, SLAO substrates were annealed at 1000°C for 2 hours in air with $LaAlO_3$ substrates covered[51]. The $La_{3-x}Sr_xNi_2O_7$ films were grown using a shuttered layer-by-layer deposition with specific deposition sequence, as outlined in prior studies[52,53]. After deposition, the films were cooled down to room temperature under the same oxidizing background pressure to minimize the oxygen vacancies.

### Ozone annealing

To achieve optimal superconductivity, the as-grown films were ex-situ annealed at 380°C for ~1 hour in a home-made ozone annealing system equipped with a commercial ozone generator (AC-2025, IN USA Inc.). During annealing, the oxygen flow rate was maintained at ~100 sccm, and the ozone generator output power was fixed at 10%. To suppress second-phase formation[16], the warming-up time to 380°C was limited to 3 minutes, and the cooling-down time (to 150°C) was limited to 6 mins. Optimal annealing conditions were systematically investigated for samples with varying Sr doping levels, as detailed in **Extended Data Fig. 5**.

### X-ray diffraction and topography characterization

X-ray diffraction data were measured using a Bruker D8 Discover diffractometer with a monochromated Cu $K_{\alpha 1}$ radiation source (λ = 1.5406 Å).

Surface topography was analyzed via atomic force microscopy using an Asylum Research MFP-3D Origin+ scanning probe microscope operated in contact mode.

**Scanning transmission electron microscopy and spectroscopy**

The cross-sectional TEM samples were prepared using a Thermo Fisher Helios 5CX Dual Beam FIB/SEM (focused ion beam/scanning electron microscope) system. A 10 nm carbon layer was deposited on the sample using Leica ACE 600, a high vacuum coating system, to protect the sample surface. Before ion etching, a protecting platinum layer was deposited on the area of interest using a low-energy electron beam and then ion beam. The thinning process was performed using gallium ion beam at 30 kV, 0.79 nA, followed by a final ion polishing at 5 kV, 0.12 nA. STEM characterization was carried out using an aberration-corrected microscope (Spectra 300, Thermo Fisher Scientific) operated at 300 kV and equipped with a Super-X spectrometer. EDS mapping was obtained with a probe current of 50 pA and a total duration of about 15 minutes to minimize electron irradiation damage. HAADF-STEM images were acquired using a 29.9 mrad convergence semi-angle and a 65-200 mrad collection angle. The iDPC images were acquired using the same convergence semi-angle and a 15-30 mrad collection angle. The lattice constants, atomic column intensity, bond length and bond angle were analyzed using the software tool Atomap[54].

**DFT calculations**

The strained $La_3Ni_2O_7$ are calculated by the density functional theory in the generalized gradient approximation implemented in the Vienna ab initio simulation package (VASP) code[55,56], in which the projected augmented wave method[57,58] and the Perdew-Burke-Ernzerhof exchange-correlation[59] are used. The plane-wave cutoff energy is 520 eV and the k-mesh is 8× 8 ×2 for the calculations for one unit cell. In all the calculations, no magnetism is included, but a Hubbard U of 4.0 eV is added to the Ni's 3d orbital, as used in the reference[1]. The crystal structure of $La_3Ni_2O_7$ is taken from the reference[60] but is then fully optimized in VASP. Our optimized lattice constants are $a$ = 5.35342 Å, $b$ = 5.40960 Å, and $c$ = 20.6652 Å, which are well consistent with the experimental ones: $a$ = 5.39283 Å, $b$ = 5.44856 Å, and $c$ = 20.51849 Å.

To simulate the strained film, we used an average in-plane constant $a_0$ based on the above optimized in-plane constants. For the in-plane strain, we change its in-plane lattice constants manually but optimize the out-of-plane one. The size of in-plane strain ε is defined as $\varepsilon = \frac{a-a_0}{a_0} \times 100\%$, where $a$ is the strained in-plane lattice constant and $a_0$ is the free-standing one.

To calculate the formation energy of the oxygen vacancy, we used $2\times 2 \times 1$ supercells (191 atoms) with one oxygen at the outer-apical, inner-apical, or planar site removed. The supercells with oxygen vacancy are also optimized with the fixed in-plane lattice constants but optimized out-of-plane ones.

**Electrical transport**

DC electrical transport measurements were performed in van der Pauw or Hall bar configurations using a Physical Property Measurement System (PPMS, Quantum Design), a TeslatronPT system (Oxford Instruments) and a home-made cryostat equipped with a 2612B System SourceMeter (Keithley).

The current was applied exclusively at temperatures below 200 K to mitigate oxygen loss at higher temperatures[15], unless otherwise specified, and fixed at 5 $\mu$A for temperature-dependent resistivity measurements and 500 $\mu$A for Hall effect characterization.

Gold electrodes were deposited using DC sputtering and patterned with shadow masks, followed by ultrasonic aluminum wire bonding. Owing to the heating effect and vacuum environment inherent to sputtering, $La_{3-x}Sr_xNi_2O_7$ films are prone to oxygen loss. Resistivity values were calculated using geometry correction factors derived from finite-element-method (FEM) simulations.

**Methods References**

## Acknowledgements

This work was supported by the National Key R&D Program of China (Grant Nos. 2022YFA1402502, 2021YFA1400400), the National Natural Science Foundation of China (Nos. 12434002) and Natural Science Foundation of Jiangsu Province (Grant Nos. BK20233001). D.J. acknowledges the National Natural Science Foundation of China (Grant Nos. 12204394), the startup grant from the Department of Applied Physics, the Hong Kong Polytechnic University (Grant No. 1-BD6B), the General Research Fund (Grant No. 15303923, 15307224) from the Hong Kong Research Grants Council. W.S. acknowledges the National Natural Science Foundation of China (Grant Nos. 123B2051). H.S. acknowledges the China National Postdoctoral Program for Innovative Talents (Grant No. BX20230152) and the China Postdoctoral Science Foundation (Grant No. 2024M751368) and the Natural Science Foundation of Jiangsu Province (Grant No. BK20241189).

## Author Contributions

Y.N. conceived the idea and directed the project. B.H. and M.W. synthesized the samples and characterized the crystalline structure with the help of S.Y., Z.M., L.H. and H.S. under the supervision of Z.G. and Y.N.. B.H., M.W., W.S., and H.Z. performed the electrical transport measurements and data analyses under the supervision of Y.N.. Y.Y. performed the STEM measurements under the supervision of D.J.. J.Z. performed DFT calculations. B.H., M.W., W.S., and Y.Y. wrote the manuscript under the supervision of J.Z., D.J. and Y.N.. All authors discussed the data and contributed to the manuscript.

## Competing interests

Authors declare no competing interests.

## Data availability

The source data presented in the figures of this study are available from the corresponding authors upon reasonable request.

# Supplementary Materials

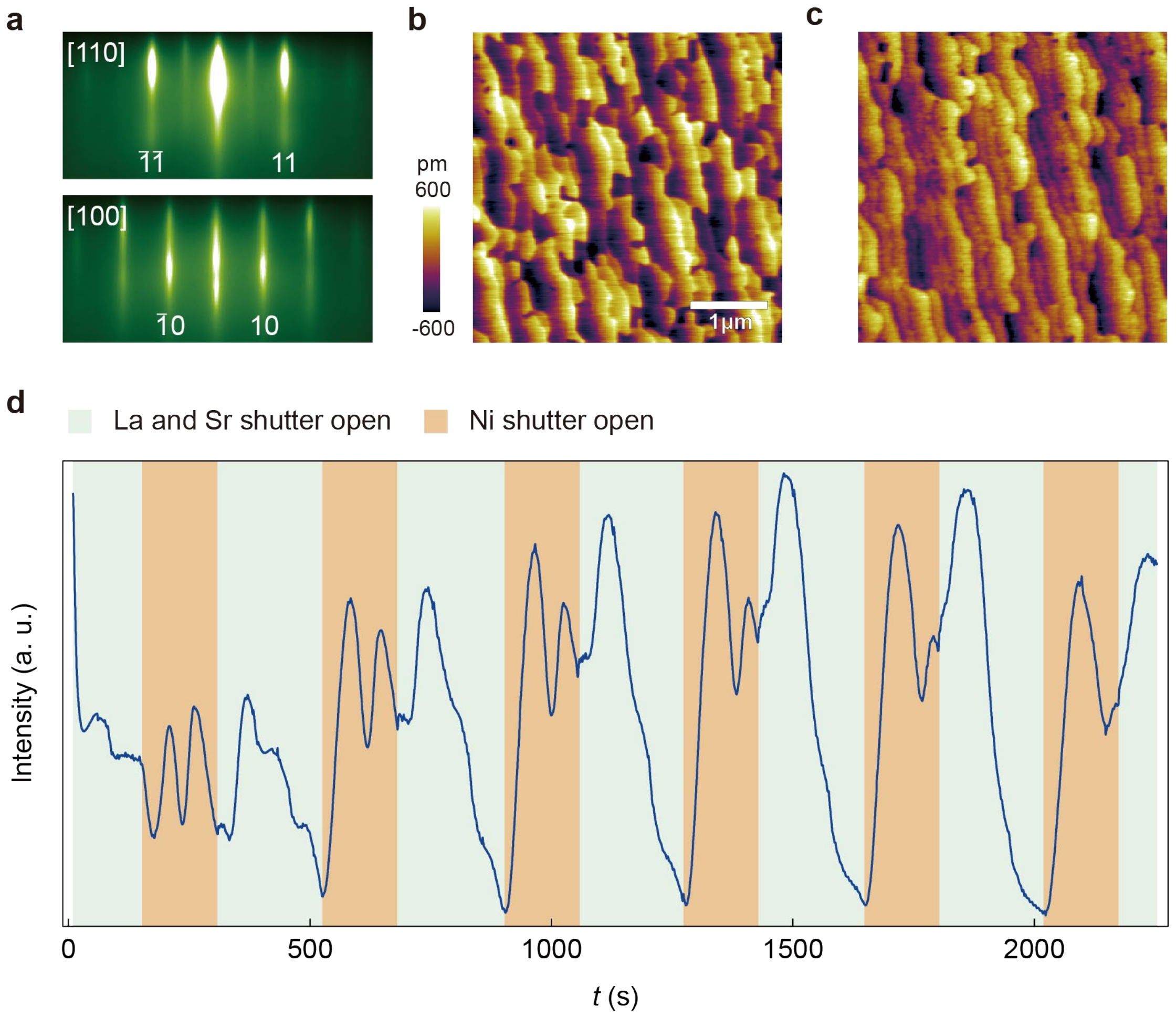

**Extended Data Figure 1: Growth and surface topography of the $La_{2.91}Sr_{0.09}Ni_2O_7$ thin film shown in Fig. 1d and Fig. 3. a**, Reflection high-energy electron diffraction (RHHED) patterns of as-grown $La_{2.91}Sr_{0.09}Ni_2O_7$ thin film taken along [110] and [100] directions. **b-c**, Atomic force microscopy of the treated $SrLaAlO_4$ substrate (**b**) before growth and the $La_{2.91}Sr_{0.09}Ni_2O_7$ thin film (**c**) after growth. **d**, The RHEED intensity oscillations of a 3-u.c.-thick $La_{2.91}Sr_{0.09}Ni_2O_7$ film grown on $SrLaAlO_4$ substrate, with the growth sequence of [(La,Sr)O] - [(La,Sr)O] - [$NiO_2$] - [$NiO_2$] - [(La,Sr)O].

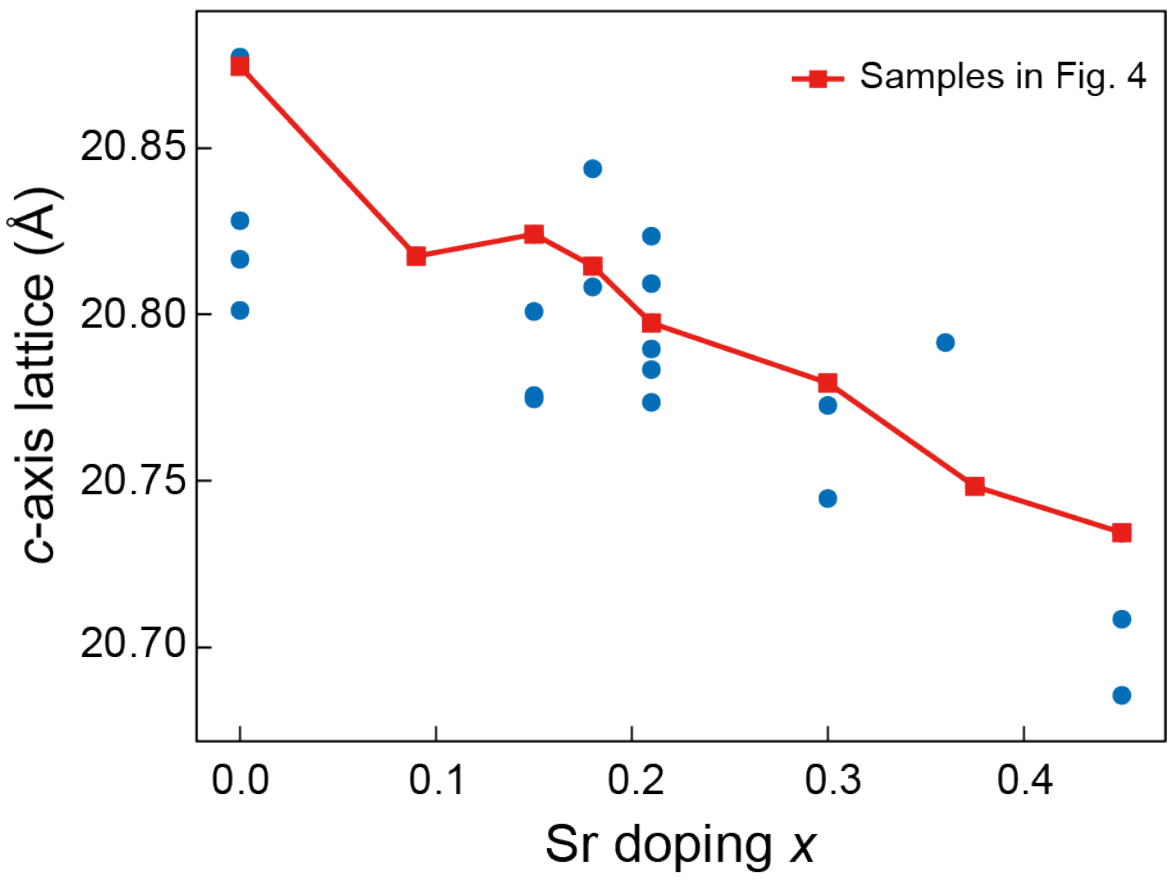

**Extended Data Figure 2: Relation between *c*-axis lattice constants and Sr doping levels *x* in $La_{3-x}Sr_xNi_2O_7$ films at room temperature.** Only samples with a lattice constant of the small standard deviation are shown. The red squares represent the same sample in Fig. 4.

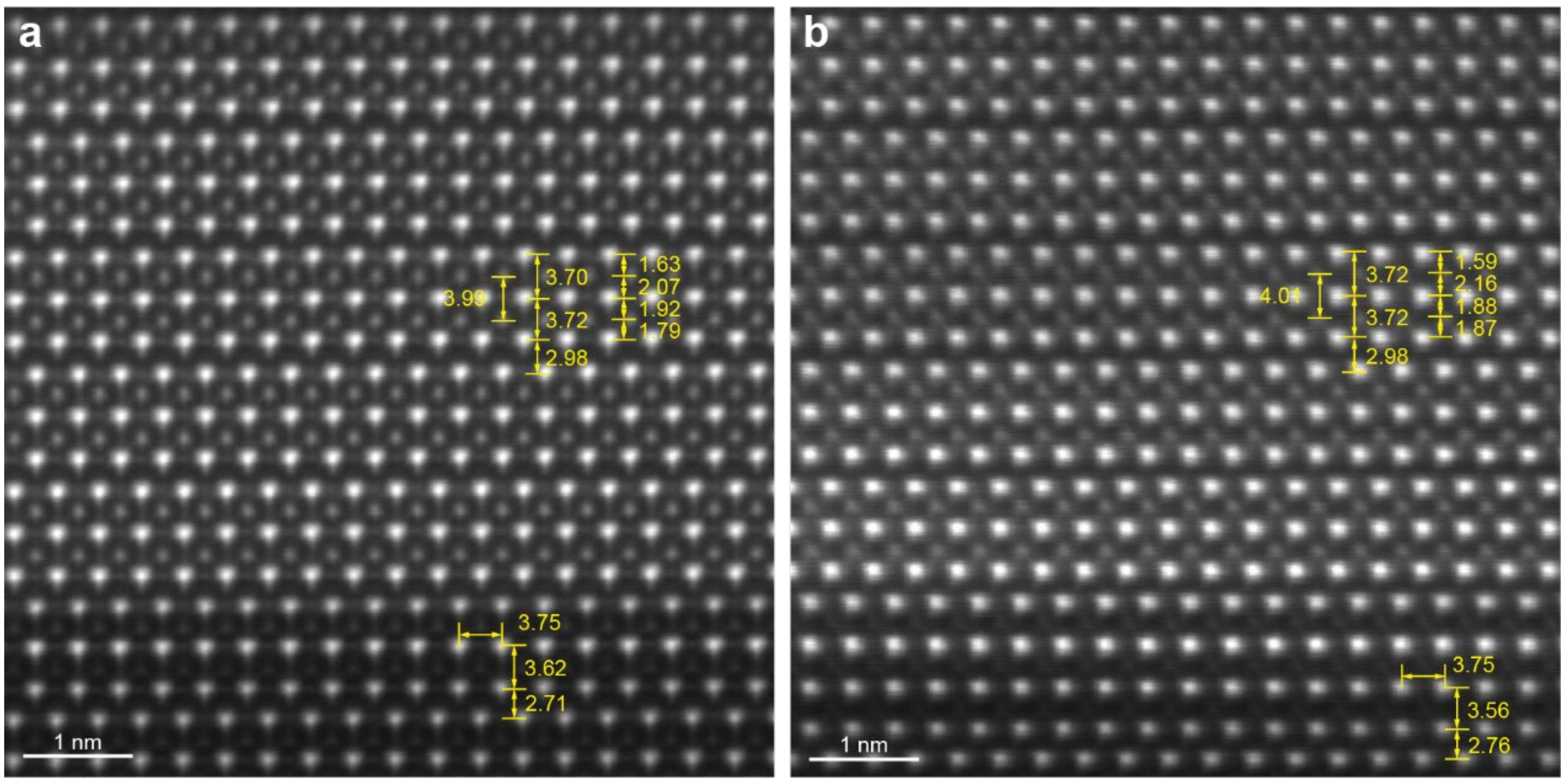

**Extended Data Figure 3:** Lattice constants of $La_{3-x}Sr_xNi_2O_7$ on SLAO in samples with $x = 0.21$ (a) and $x = 0$ (b).

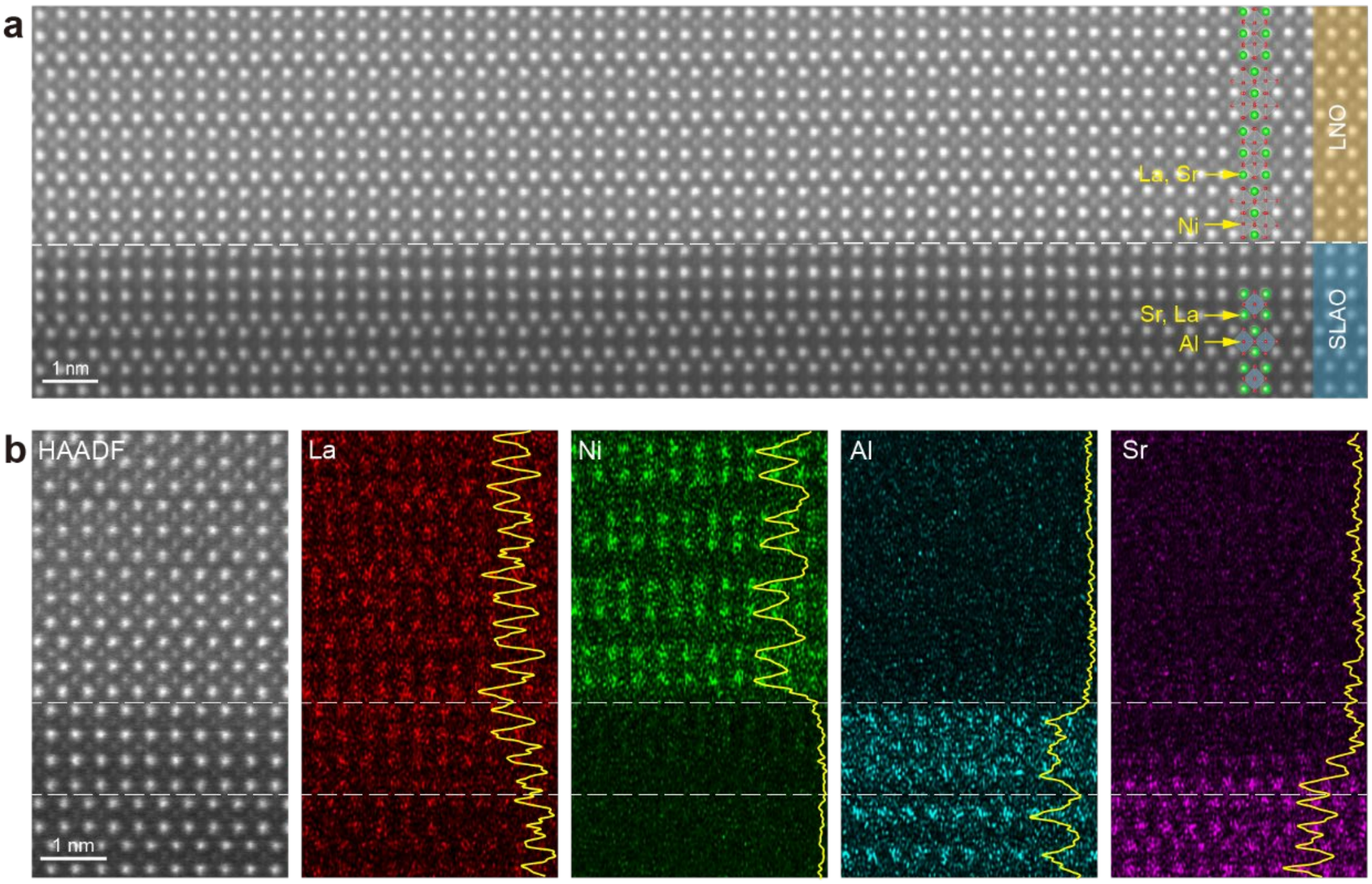

**Extended Data Figure 4: Scanning transmission electron microscopy (STEM) characterizations of a superconducting $La_3Ni_2O_7$ film. a**, High-angle annular dark-field (HAADF) image with a large field-of-view of a 3-u.c.-thick $La_3Ni_2O_7$ film grown on a $SrLaAlO_4$ substrate. Dash lines represent the interface between the film and the $SrLaAlO_4$ substrate. **b**, HAADF image and atomic-resolution energy-dispersive X-ray spectroscopy (EDS) elemental maps (La, Ni, Al, Sr) of the same region. The yellow curves represent the profiles of atomic row-integrated elements intensity. The area between the dashed lines indicates the surface reconstruction region of the $SrLaAlO_4$ substrate.

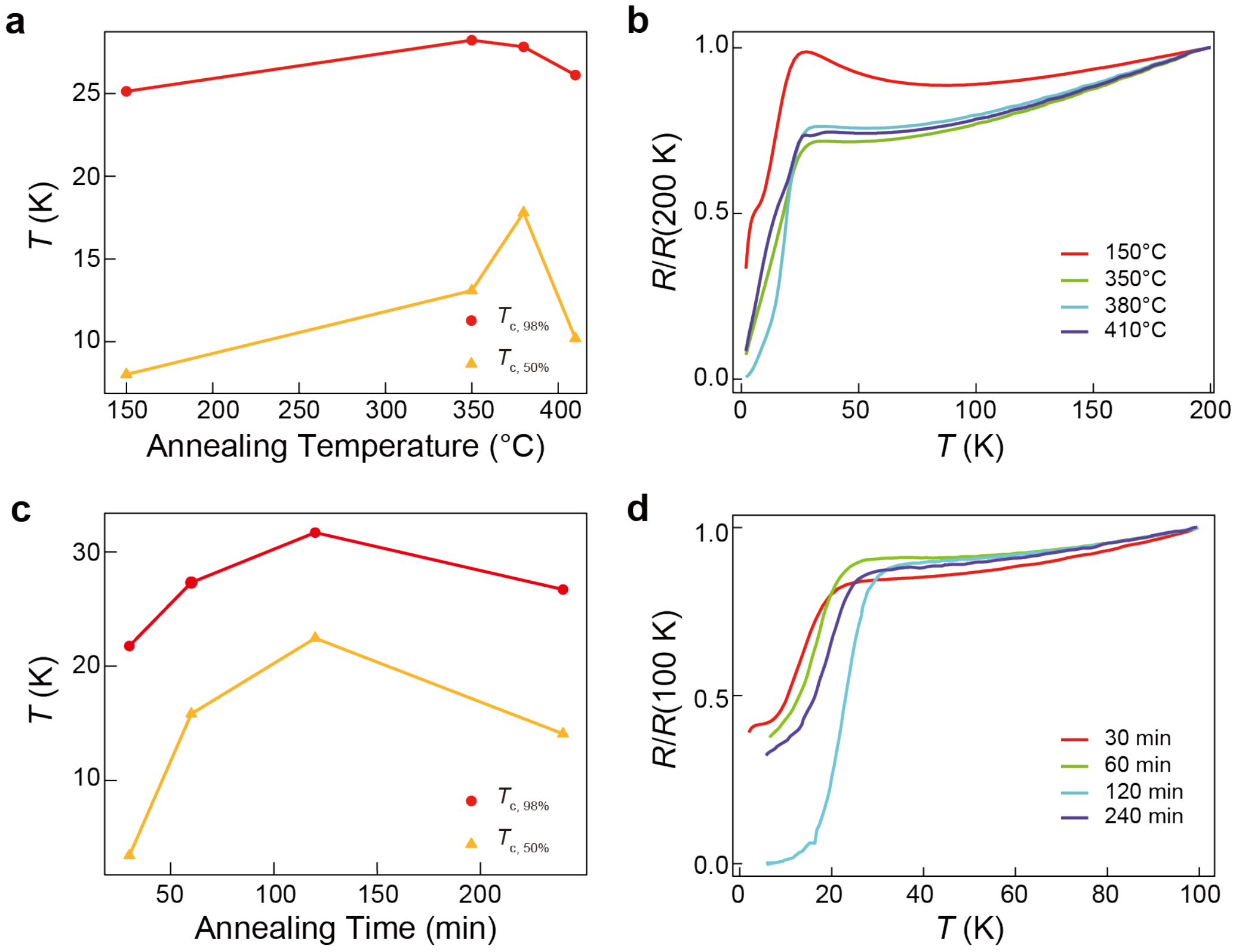

**Extended Data Figure 5: The optimalization of annealing conditions for $La_{2.79}Sr_{0.21}Ni_2O_7$ films. a**, The $T_{c,98\%}$ and $T_{c,50\%}$ of the $La_{2.79}Sr_{0.21}Ni_2O_7$ films annealed under different temperature, and related temperature-dependence of resistance (normalized at 200 K) (**b**). **c**, The $T_{c,98\%}$ and $T_{c,50\%}$ of the $La_{2.79}Sr_{0.21}Ni_2O_7$ films annealed at 380 °C with different time, and related temperature-dependence of resistance (normalized at 100 K) (**d**). Each series of experiments were conducted on separate pieces from the same sample with a fixed output power of the ozone generator.

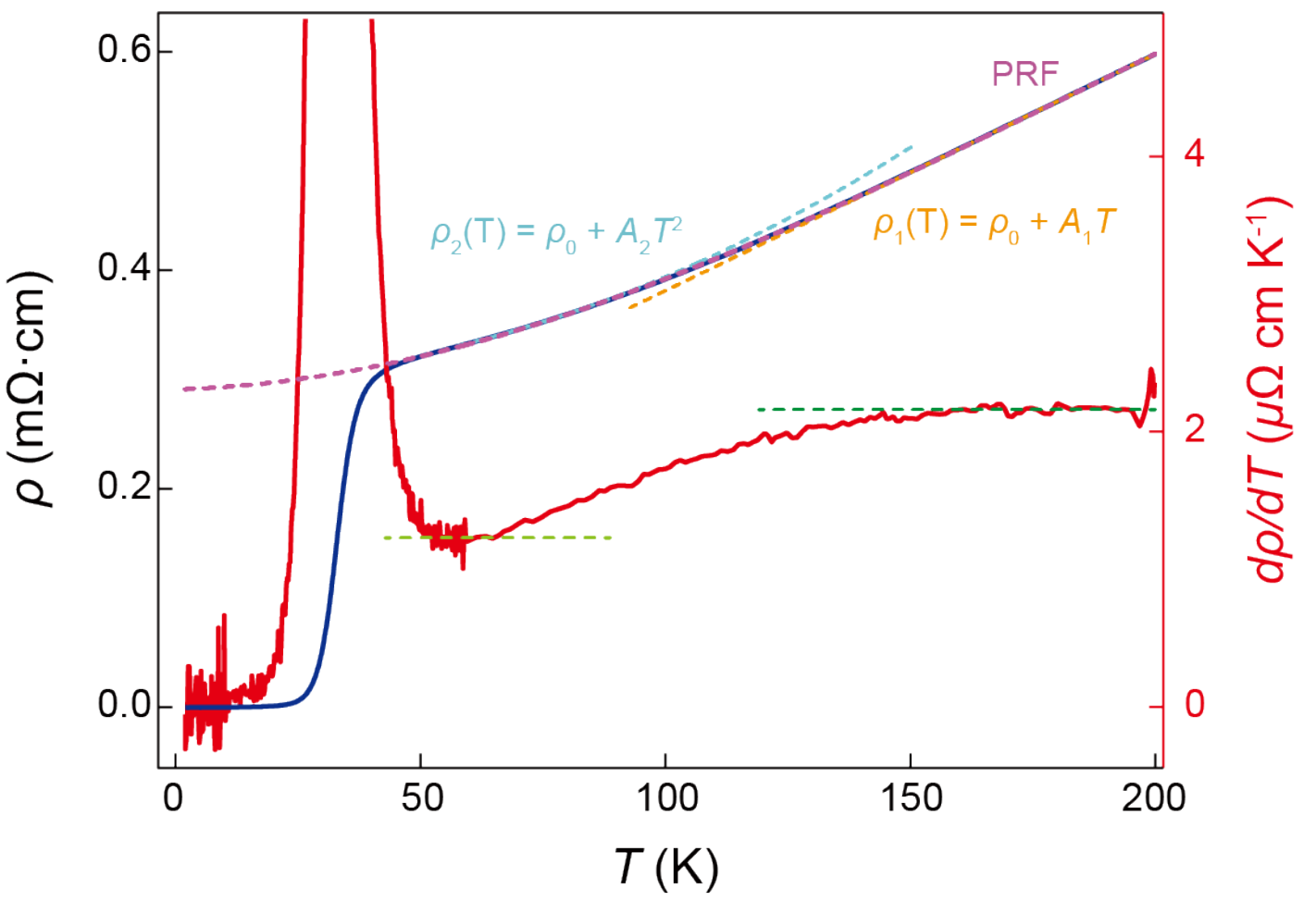

**Extended Data Figure 6:** Analysis of RT curve of the $La_{2.91}Sr_{0.09}Ni_2O_7$ film shown in **Figure 1d**. Red curve is the temperature-dependent deviation of resistivity.

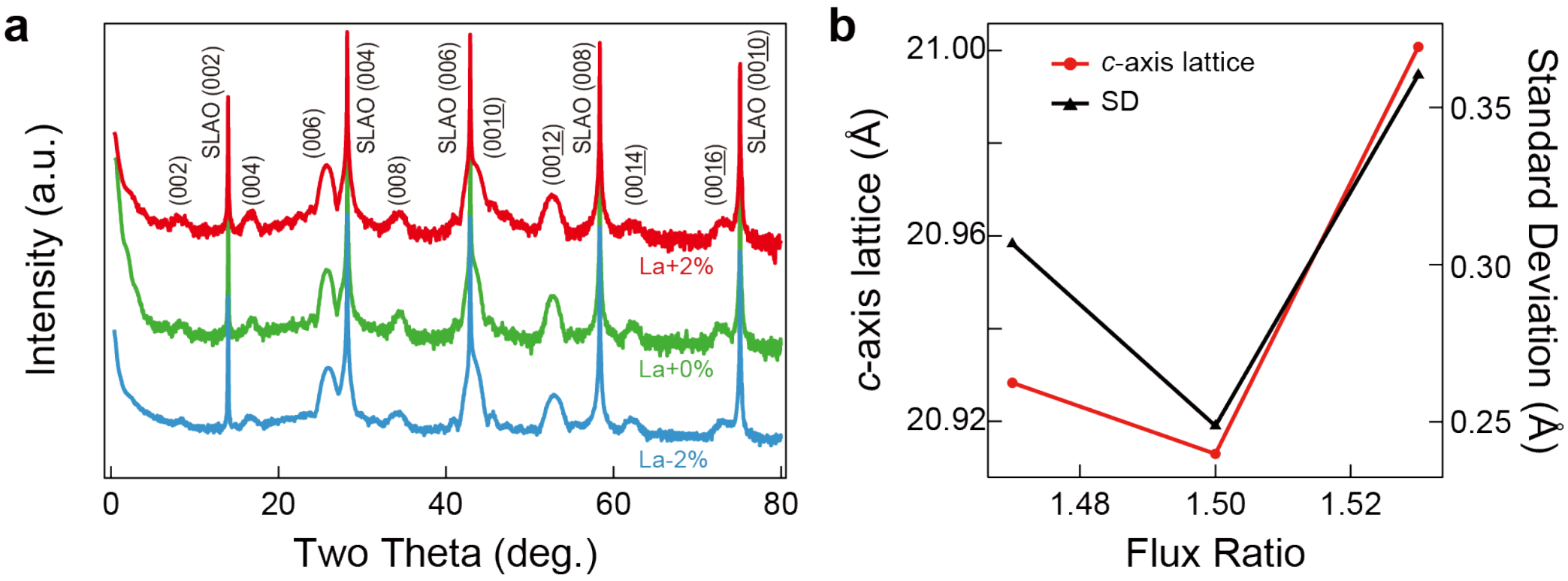

**Extended Data Figure 7: Structural characterization of $La_{2.79}Sr_{0.21}Ni_2O_7$ films with varying La content. a**, XRD of 2.5-u.c.-thick $La_{2.79}Sr_{0.21}Ni_2O_7$ films on $SrLaAlO_4$. **b**, The average and the standard deviation of the *c*-axis lattice constants. The optimal flux ratio for each Sr doping level $x$ is calibrated using XRD measurements, by identifying the sample with a lattice constant of the smallest standard deviation. The change of La content is controlled via La shutter time.

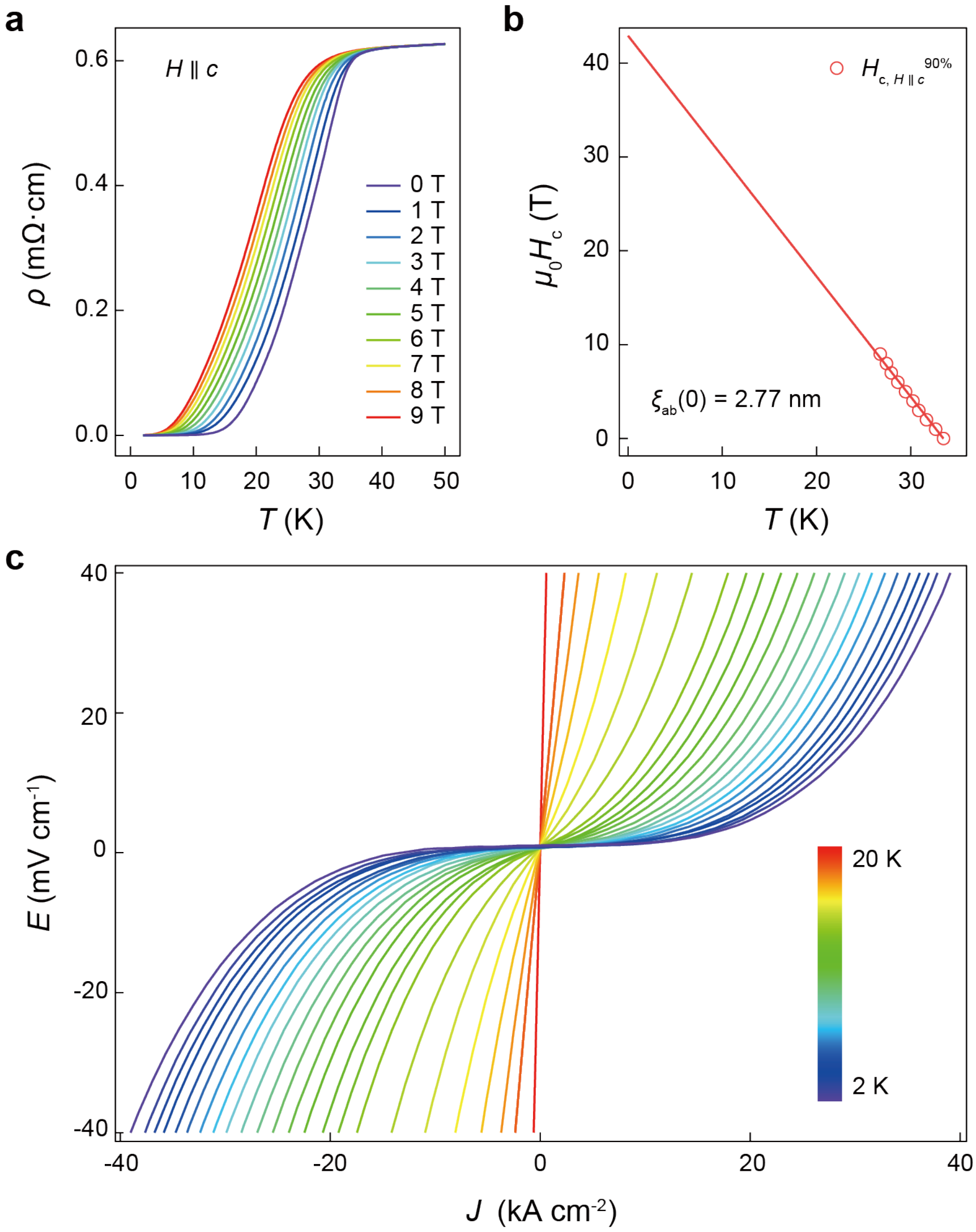

**Extended Data Figure 8: Transport properties of a $La_{2.79}Sr_{0.21}Ni_2O_7$ thin film. a,** $\rho(T)$ curves of $La_{2.79}Sr_{0.21}Ni_2O_7$ thin film under various magnetic fields applied perpendicular to the *a*-*b* plane of the film. **b,** Perpendicular upper critical fields extracted by the $T_{c,90\%}$ values (open circles). Solid lines represent Ginzburg-Landau fitting results. **c,** Electric field ($E$) versus current density ($J$) curves measured at 2-20 K.

| In-plane biaxial strain | Formation energy of oxygen vacancies (eV) | | |
|---|---|---|---|
| | Outer apical | Planar | Inner apical |
| 0% | 0 | -1.086 | **-1.554** |
| -2% | 0 | -2.025 | -2.040 |
| -4% | 0 | **-2.110** | -1.931 |

**Extended Data Table 1: Calculated formation energies for three types of oxygen vacancies under varying in-plane biaxial strains.** For each strain condition, the formation energy of the outer apical oxygen vacancy is defined as the reference (zero) energy. Without external strain, the inner apical oxygen vacancy exhibits the lowest formation energy. Under -4% compressive in-plane biaxial strain, oxygen vacancies form more readily at planar sites.